\documentclass[11pt,journal,onecolumn]{IEEEtran}

\usepackage{ifpdf}
\usepackage{graphics}
\usepackage{color}
\usepackage{epsfig}
\usepackage{mathptmx}
\usepackage{times}
\usepackage{algpseudocode}
\usepackage{algorithmicx}
\usepackage{amssymb}
\usepackage{amsmath}
\usepackage{amsfonts}
\usepackage{bm}
\usepackage{graphicx}
\usepackage{psfrag}
\usepackage{url}
\usepackage{multirow}
\usepackage{mathrsfs}
\usepackage{enumerate}
\usepackage{midfloat}
\usepackage{pdfpages}
\usepackage[latin1]{inputenc}

\newtheorem{proposition}{Proposition}
\newtheorem{remark}{Remark}

\newcommand{\be}{\begin{equation}}
\newcommand{\ee}{\end{equation}}
\newcommand{\ben}{\begin{equation*}}
\newcommand{\een}{\end{equation*}}
\newcommand{\ba}{\begin{array}}
\newcommand{\ea}{\end{array}}

\begin{document}

\title{MAP moving horizon estimation for threshold measurements with application to field monitoring}

\author{
\IEEEauthorblockN{G. Battistelli$^{a}$, L. Chisci$^{a}$, N. Forti$^{a,b}$, S. Gherardini$^{c}$} \\
\IEEEauthorblockA{$^{a}$Universit\`a  di Firenze, Dipartimento di Ingegneria dell'Informazione (DINFO), Via di Santa Marta 3, 50139 Firenze, Italy. \\
$^{b}$ Research Department, NATO Science and Technology Organisation - Centre for Maritime Research and Experimentation (STO - CMRE), La Spezia, Italy. \\
$^{c}$ Dipartimento di Fisica e Astronomia \& LENS, Universit\`a di Firenze, via G. Sansone 1, 50019 Sesto Fiorentino, Italy and Istituto di Fisica Nucleare (INFN), Sezione di Firenze. \\
}}

\maketitle

\begin{abstract}
The paper deals with state estimation of a spatially distributed system given noisy measurements from pointwise-in-time-and-space threshold sensors spread over the spatial domain of interest. A \textit{Maximum A posteriori Probability} (MAP) approach is undertaken and a \textit{Moving Horizon} (MH) approximation of the MAP cost-function is adopted. It is proved that, under system linearity and log-concavity of the noise probability density functions, the proposed MH-MAP state estimator amounts to the solution, at each sampling interval, of a convex optimization problem. Moreover, a suitable centralized solution for large-scale systems is proposed with a substantial decrease of the computational complexity. The latter algorithm is shown to be feasible for the state estimation of spatially-dependent dynamic fields described by  \textit{Partial Differential Equations} (PDE) via the use of the \textit{Finite Element} (FE) spatial discretization method. A simulation case-study concerning estimation of a diffusion field is presented in order to demonstrate the effectiveness of the proposed approach. Quite remarkably, the numerical tests exhibit a \textit{noise-assisted} behavior of the proposed approach in that the estimation accuracy results optimal in the presence of measurement noise with non-null variance.
\end{abstract}


\section{Introduction}

Threshold sensors, which provide a binary output just indicating whether the noisy measurement of the sensed variable falls below or above a given threshold, are widely used for monitoring and control
\cite{Wang1}-\cite{SelvaratnamIEEE2017}.
The motivation is that by a multitude of low-cost and low-resolution sensing devices it is possible to attain the same estimation accuracy that a fewer (even a single one) high-cost and high-resolution ones could provide, but with significant practical advantages in terms of ease of sensor deployment and minimization of communication requirements. Since a threshold measurement just conveys the least possible amount (i.e., a single bit) of information, while implying communication bandwidth savings and consequent improved energy efficiency, it becomes of paramount importance to fully exploit the little available information by means of smart estimation algorithms. Over the last two decades, interesting work has been devoted to system identification, \cite{Wang1,Wang2} parameter
\cite{Ristic}-\cite{Ribeiro2}
and state estimation,
\cite{Irr-sampling}-\cite{Capponi}
with a specific focus on source localisation, \cite{RisticAE2016,SelvaratnamIEEE2017} using threshold measurements according to either a deterministic
\cite{Wang1}-\cite{Bai}
or a probabilistic
\cite{Ristic}-\cite{SelvaratnamIEEE2017}
approach.

In a deterministic setting, \cite{Irr-sampling,Gherardini2} the information provided by a threshold sensor is mainly associated to the switching instants, corresponding to discontinuities of the threshold signal. In fact, a switch at the current discrete time implies, by continuity, the existence of a continuous time instant, within the latest sampling interval, such that the continuous output crosses the threshold at that time. As shown in previous work, \cite{Gherardini2} additional information can also be exploited in the non-switching sampling instants by penalizing values of the estimated variable such that the corresponding predicted measurement is on the opposite side, with respect to a threshold sensor reading, far away from the threshold. Nevertheless, it is clear that there is no or very little information available for estimation purposes whenever no or very few threshold sensor switchings occur. Hence, a possible way to achieve high estimation accuracy is to have many threshold sensors sensing the same variable with different thresholds as this would clearly increase the number of switchings, actually emulating, when the number of sensors grows to infinity, a single continuous-valued (analog) measurement.

Conversely, following a probabilistic approach, each threshold sensor can be characterized in terms of the probability that its binary output takes the two possible outcomes, in relation to the dynamical evolution of the state of the monitored system. In this way, also a threshold sensor always provides an informative contribution.
Furthermore, it is worth noting how, in presence of a sufficiently large number of threshold sensors, such a stochastic approach can also be more advantageous in case the outcomes from the sensors are noisy. Indeed, as it will be shown via simulation experiments, there is a non-null value of the measurement noise variance, for which optimal state estimation accuracy from threshold measurements is obtained. This, referred to hereafter as \textit{noise-assisted} paradigm, \cite{ACC_binary} is in sharp contrast with what happens in the case of linear sensors, and depends on the strong nonlinearity of threshold sensors.

From the above discussed \textit{noise-assisted} paradigm, this paper develops a novel approach to recursive estimation of the state of a discrete-time dynamical system given threshold measurements. The proposed approach relies on a \textit{Moving-Horizon} (MH) approximation 
\cite{Morari}-\cite{SchneiderIEEE2017}
of the \textit{Maximum A-posteriori Probability} (MAP) estimation \cite{Delgado} and extends previous work \cite{Ristic,Wong} concerning parameter estimation to recursive state estimation. A further contribution is to show that, for a linear system and log-concave probability distributions, the optimization problem arising from the MH-MAP formulation turns out to be convex and, hence, practically feasible for real-time implementation.

The present paper significantly expands our preliminary work \cite{ACC_binary} by considering a more general class of probability distributions,
by providing a mathematical proof of the above stated convexity result and, most importantly, by proposing a novel efficient MH-MAP filter for field estimation of large-scale systems. The paper is structured as follows. Section 2 formulates \textit{Maximum A posteriori Probability} (MAP) state estimation with threshold measurements,
relying on the probabilistic description of the amount of information provided by each threshold sensor. Section 3 introduces its \textit{moving-horizon} (MH) approximation, referred to as MH-MAP estimator, and analyzes the properties of the resulting optimization problem. Section 4 presents the finite element approximation for the estimation of diffusion fields from threshold pointwise-in-space-and-time field measurements. Due to the large-scale of the dynamical system resulting from spatial discretization of the field dynamics, a fast MH-MAP filter for field estimation of large-scale systems is proposed in Section 5. Section 6 shows simulation results relative to the dynamic field estimation case-study, while Section 7 ends  the paper with some discussion and perspectives for future work.

\section{Maximum a posteriori state estimation with threshold sensors}

Let us consider the problem of recursively estimating the state of the discrete-time nonlinear dynamical system
\begin{equation}\label{1}
\begin{array}{rcl}
x_{k+1} &=& f(x_{k},u_{k})+ w_{k} \\
z_{k}^{(i)} &=& h^{(i)}(x_{k})+ v_{k}^{(i)},\hspace{3mm}i=1,\ldots,l
\end{array}
\end{equation}
from a set of measurements provided by threshold sensors
\begin{equation} \label{2}
\begin{array}{rclcl}
y_{k}^{(i)} & = & g^{(i)}(z_{k}^{(i)}) & = &  \left\{
\begin{array}{ll} 1, & \mbox{if }  z_{k}^{(i)} \geq \tau^{(i)} \\
                            0,   & \mbox{if }  z_{k}^{(i)}  < \tau^{(i)}
\end{array}
\right.
\end{array}
\end{equation}
where $x_k \in \mathbb{R}^n$ is the state to be estimated, $u_k \in \mathbb{R}^m$ is a known input, and $\tau^{(i)}$ is the threshold of the $i-$th sensor.
The vector $w_{k}\in \mathbb{R}^n$ is an additive disturbance affecting the system dynamics while $v^{(i)}_k$ is the measurement noise of sensor $i$.
Notice from (\ref{1})-(\ref{2}) that sensor $i$ produces a threshold measurement $y^{(i)}_{k}\in\{0,1\}$ depending on whether the noisy system output $z^{(i)}_{k}$ is below or above the threshold $\tau^{(i)}$.
We define, for the sake of simplicity,
\begin{equation}
z_k = {\rm col} \left ( z_k^{(1)} , \ldots, z_k^{(l)} \right )  \, , \quad y_k = {\rm col} \left ( y_k^{(1)} , \ldots, y_k^{(l)} \right ) \, , \quad
v_k = {\rm col} \left ( v_k^{(1)} , \ldots, v_k^{(l)} \right )  \, .
\end{equation}

Let $\mathcal{N}(\mu,\Sigma)$ denote the normal distribution with mean $\mu$ and variance $\Sigma$. Then the statistical behavior of the system is modelled according to
the following assumption.
\begin{enumerate}[\bf {A}1]
\item
The initial state $x_{0}$ and disturbance $w_{k}$ are normally-distributed random vectors
\begin{equation}
x_{0}\sim\mathcal{N}(\overline{x}_{0},P^{-1}),\hspace{2mm}w_{k}\sim\mathcal{N}(0,G^{-1})
\label{prob}
\end{equation}
where $\mathbb{E}[w_{j}w_{k}']=0$ if $j\neq k$ and $\mathbb{E}[w_{j}x_{0}']=0$ for any $j$. Further, the measurement noises $v_k^{i}$ of the sensors are mutually independent as well as independent from the initial state and disturbance.
\end{enumerate}

According to the available probabilistic description, hereafter the problem of state estimation from threshold measurements is recast into a Bayesian framework exploiting a MAP estimation approach. As discussed in the introduction, each threshold measurement $y^{(i)}_k$ provides intrinsically relevant probabilistic information on the state $x_k$.
Such information can be effectively exploited by introducing the likelihood functions $p(y^{(i)}_{k}|x_{k})$ of the $i-$th threshold sensor.
To this end, let us observe that each threshold measurement $y^{(i)}_{k}$ is a Bernoulli random variable such that, for any threshold sensor $i$ and any time instant $k$, the likelihood function $p(y^{(i)}_{k}|x_{k})$ is given by \cite{Ristic,Ribeiro1}
\begin{equation}
p(y^{(i)}_{k}|x_{k})~=~p(y^{(i)}_{k}=1|x_{k})^{y^{(i)}_{k}} ~p(y^{(i)}_{k}=0|x_{k})^{1-y^{(i)}_{k}}
\end{equation}
where
\begin{equation}\label{8}
p(y^{(i)}_{k}=0|x_{k}) = F^{(i)}(\tau^{(i)}-h^{(i)}(x_{k}))
\end{equation}
and $p(y^{(i)}_{k}=1 | x_{k})=1-p(y^{(i)}_{k}=0|x_{k}) \triangleq 1 - F^{(i)}(\tau^{(i)}-h^{(i)}(x_{k}))$.
In particular, $F^{(i)}(\tau^{(i)}-h^{(i)}(x_{k})) = prob( v_k^{(i)} \leq \tau^{(i)} - h^{(i)}(x_k))$ is the
\textit{Cumulative Distribution Function} (CDF) of the random variable $v_k^{(i)}$ evaluated at $\tau^{(i)}-h^{(i)}(x_{k})$.
For example, when the measurement noise is normally distributed
 $v^{(i)}_{k} \sim \mathcal{N}(0,r^{(i)})$, the conditional probability $p(y^{(i)}_{k}=1|x_{k}) = 1-F^{(i)}(\tau^{(i)}-h^{(i)}(x_{k})$ can be written in terms of a $Q$-function, describing the tail probability of a standard normal probability distribution \cite{Wim2006}\begin{equation}
p(y^{(i)}_{k}=1 | x_{k}) = \frac{1}{\sqrt{2\pi r^{(i)}}}\int_{\tau^{(i)}-h^{(i)}(x)}^{\infty}\exp\left(-\frac{u^{2}}{2r^{(i)}}\right) du
= Q \left(\frac{\tau^{(i)}-h^{(i)}(x)}{\sqrt{r^{(i)}}}\right) \, . \label{eq7}
\end{equation}

Let us now denote by $Y_{k} \triangleq {\rm col} (y_0,\ldots,y_{k} )$ the vector of all threshold measurements collected up to time $k$ and by  $X_{k}\triangleq {\rm col} (x_{0},\ldots,x_{k} )$ the vector of the state trajectory. Further, let us denote by $\hat{X}_{k|k} \triangleq {\rm col} ( \hat{x}_{0|k},\ldots,\hat{x}_{k|k} )$ the estimate of $X_{k}$ at time $k$. Then, at each time instant $k$, given the a posteriori probability $p(X_{k}|Y_{k})$, the estimate of the state trajectory can be obtained by solving the following MAP estimation problem:
\begin{equation}
\hat{X}_{k|k}  = \text{arg}\max_{X_{k}}p(X_{k}|Y_{k})
=\text{arg} \min_{X_{k}}  \left\{ - \ln p(X_{k}|Y_{k}) \right\} \label{12}.
\end{equation}
From the Bayes rule
\begin{equation}
p(X_{k}|Y_{k}) ~\propto ~p(Y_{k}|X_{k})~p(X_{k}),
\end{equation}
where $p(Y_{k}|X_{k})$ is the likelihood function of the threshold measurement vector $Y_{k}$, and
\begin{equation}
p(X_{k}) = \prod_{j=0}^{k-1}p(x_{k-j}|x_{k-j-1},\ldots,x_{0})~p(x_{0}) = \prod_{j=0}^{k-1}p(x_{k-j}|x_{k-j-1})~p(x_{0})
\end{equation}
thanks to the Markov property of the system state.
Furthermore, in view of assumption A1, we have
\begin{eqnarray}
p(x_0) &\propto & \exp\left(-\frac{1}{2}\|x_{0}-\overline{x}_0 \|^{2}_{P}\right)\label{15} \\
p(x_{k+1}|x_{k}) &\propto& \exp\left(-\frac{1}{2}\|x_{k+1}-f(x_{k},u_{k})\|^{2}_{G}\right)\label{16},
\end{eqnarray}
so that
\begin{equation}
p(X_{k}) = \exp\left(-\frac{1}{2}\left[\|x_{0}-\overline{x}_{0}\|^{2}_{P}+\sum_{j=0}^{k}\|x_{j+1}-f(x_{j},u_{j})\|^{2}_{G}\right]\right).
\end{equation}
In view of the mutual independence of the measurement noises,
the likelihood function $p(Y_{k}|X_{k})$ can be written as
\begin{equation}
p(Y_{k}|X_{k})=\displaystyle{\prod_{j=0}^{k}}p(y_{j}|x_{j})=\prod_{j=0}^{k}~\prod_{i=1}^{l}p(y_{j}^{(i)}|x_{j})
=\displaystyle{\prod_{j=0}^{k}~\prod_{i=1}^{l}} F^{(i)}(\tau^{(i)}-h^{(i)}(x_{j}))^{1-y_{j}^{(i)}} ~ \left[ 1 - F^{(i)}(\tau^{(i)}-h^{(i)}(x_{j})) \right]^{y_{j}^{(i)}} .
\end{equation}

In conclusion, the log-likelihood function, natural logarithm of the likelihood function, is given by
\begin{equation}
\ln p(Y_{k}|X_{k})=\displaystyle{\sum_{j=0}^{k}~\sum_{i=1}^{l}}\left\{ (1-y_{j}^{(i)}) \, \ln F^{(i)}(\tau^{(i)}-h^{(i)}(x_{j}))
+ y_{j}^{(i)} \, \ln \left[ 1 - F^{(i)} (\tau^{(i)}-h^{(i)}(x_{j}))  \right] \right\},
\end{equation}
and the cost function $J_k(X_k) =-\ln p(Y_{k}|X_{k})-\ln p(X_{k})$ to be minimized in the MAP estimation problem (\ref{12}), up to the constant term $p(Y_k)$, turns out to be
\begin{eqnarray}\label{18}
J_{k} (X_k) &=& \|x_{0}-\overline{x}_{0}\|^{2}_{P}+\displaystyle{\sum_{j=0}^{k}}\|x_{j+1}-f(x_{j},u_{j})\|^{2}_{G}\nonumber \\
&-&\displaystyle{\sum_{j=0}^{k}~\sum_{i=1}^{l}}\left\{ (1-y_{j}^{(i)}) \ln F^{(i)}(\tau^{(i)}-h^{(i)}(x_{j}))+y_{j}^{(i)} \ln \left[ 1 - F^{(i)}(\tau^{(i)}-h^{(i)}(x_{j})) \right] \right\} \, ,
\end{eqnarray}
which is defined for all vectors $X_k$ such that the arguments of the logarithms are different from zero.
Unfortunately, a closed-form expression for the global minimum of (\ref{18}) does not exist and, hence, the optimal MAP estimate $\hat X_{k|k}$ has to be determined by resorting to some numerical optimization routine.
In this respect, the main drawback is that the number of optimization variables grows linearly with time, since the vector $X_k$ has size $(k+1) \, n$.
As a consequence, as time $k$ grows the solution of the full information MAP state estimation problem (\ref{12}) becomes eventually unfeasible, and some approximation has to be introduced.

\section{Moving-horizon approximation}
\label{section_MH}

In this section, we propose an approximate method, based on the MHE approach, 
\cite{Morari}-\cite{SchneiderIEEE2017}
to solve the MAP state estimation problem.
To this end, let us introduce the sliding window $\mathfrak W_k = \{k-N, k-N+1, \ldots, k\}$, so that the goal of the estimation problem becomes that to find an estimate of the partial state trajectory $X_{k-N:k} \triangleq {\rm col} ( x_{k-N},\ldots,x_{k} )$ by using the information available in $\mathfrak W_k$.
Therefore, in place of the full information cost $J_k (X_k)$, at each time instant $k$ the minimization of the following \textit{moving-horizon cost} is considered:
\begin{eqnarray}\label{MH}
J_{k}^{\rm MH} (X_{k-N:k}) &=& \Gamma_{k-N} (x_{k-N}) + \displaystyle{\sum_{j=k-N}^{k}}\|x_{j+1}-f(x_{j},u_{j})\|^{2}_{G}\nonumber \\
&-&\displaystyle{\sum_{j=k-N}^{k}~\sum_{i=1}^{l}}\left\{ (1-y_{j}^{(i)}) \ln F^{(i)}(\tau^{(i)}-h^{(i)}(x_{j}))+ y_{j}^{(i)} \ln \left[ 1 -F^{(i)}(\tau^{(i)}-h^{(i)}(x_{j})) \right] \right\},
\end{eqnarray}
where the non-negative initial penalty function $\Gamma_{k-N} (x_{k-N}) $, known in the MHE literature as \textit{arrival cost}, \cite{Rao2,CDC} is introduced so as to summarize the past data $y_0, \ldots, y_{k-N-1}$ not explicitly accounted for in the objective function. The form of the arrival cost plays an important role in the behavior and performance of the overall estimation scheme.
While in principle $\Gamma_{k-N} (x_{k-N})$ could be chosen so that minimization of (\ref{MH}) yields the same estimate that would be obtained by minimizing (\ref{18}), an algebraic expression for such a true arrival cost seldom exists, even when the sensors provide continuous (non-threshold) measurements. \cite{Rao2}
Hence, some approximation must be used. With this respect, a common choice, \cite{NLMHE,CDC} also followed in the present work, consists of assigning to the arrival cost a fixed structure penalizing the distance of the state $x_{k-N}$ at the beginning of the sliding window from some prediction $\overline{x}_{k-N} $ computed at the previous time instant, thus making the estimation scheme recursive. A natural choice is then a quadratic arrival cost of the form
\begin{equation}\label{eq:arrival}
\Gamma_{k-N} (x_{k-N}) = \|x_{k-N}-\overline{x}_{k-N}\|^{2}_{\Psi} \, ,
\end{equation}
which, from the Bayesian point of view, corresponds to approximating the PDF of the state $x_{k-N}$ conditioned to all the measurements collected up
to time $k-1$ with a Gaussian having mean $\overline{x}_{k-N}$ and covariance $\Psi^{-1}$. As for the choice of the weight matrix $\Psi$,
in the case of continuous measurements it has been shown \cite{NLMHE,CDC} that stability of the estimation error dynamics can be ensured provided that $\Psi$
is not too large (so as to avoid an overconfidence on the available estimates). Recently, \cite{Gherardini2} similar results have been proven to hold also in the case of threshold sensors in a deterministic context. In practice, $\Psi$ can be seen as a design parameter which has to be tuned by pursuing a suitable trade-off between such stability considerations and the necessity of not neglecting the already available information (since in the limit for $\Psi$ going to zero the approach becomes a finite memory one).

It is worth noting that when the PDFs of the measurement noises are always strictly positive, like in the case of the normal distribution, then
the cost function (\ref{MH}) is well defined for any $X_{k-N:k} \in \mathbb R^{n (N+1)}$ because the arguments of the logarithms are always strictly positive. Otherwise,
in the minimization of the cost function (\ref{MH}) only the vectors $X_{k-N:k}$ for which the arguments of the logarithms are strictly positive have to be taken into account. Hereafter, we denote the set of such vectors
as $\mathbb X_k \subseteq  \mathbb R^{n (N+1)}$. Summing up, at any time instant $k = N,N+1,\ldots$, the following problem has to be solved. \\ \\ 
\textbf{Problem $E_{k}$:} Given the prediction $\overline{x}_{k-N}$, the input sequence $\{ u_{k-N}, \ldots, u_{k-1} \}$ and the measurement sequences $\{ y^{(i)}_{k-N}, \ldots, y^{(i)}_k ; \, i = 1, \ldots, l \}$, find the optimal estimates
$\hat X_{k-N:k} = {\rm col} ( \hat{x}_{k-N|k},\ldots,\hat{x}_{k|k} )$ that minimize the cost function (\ref{MH}) with arrival cost (\ref{eq:arrival}) under the constraint $X_{k-N:k} \in \mathbb X_k$. \\
\begin{remark}
In order to propagate the estimation procedure from Problem $E_{k-1}$ to Problem $E_{k}$, the prediction
$\overline{x}_{k-N}$ is set equal to the value of the estimate of $x_{k-N}$ made at time instant $k-1$, i.e., $\overline{x}_{k-N} = \hat{x}_{k-N|k-1}$.
Clearly, the recursion is initialized with the a priori expected value $\overline{x}_0$ of the initial state vector. 
\end{remark}\vspace{0.15cm}
In general, solving Problem $E_{k}$ entails the solution of a non-trivial optimization problem. However, when both equations (\ref{1}) and (\ref{2}) are linear, the resulting optimization problem turns out to be convex
for a large class of measurement noise distributions so that standard optimization routines can be used in order to find the global minimum.
To see this, let us consider the following assumptions.
\vspace{.15cm}
\begin{enumerate}[\bf {A}2]
\item The functions $f(\cdot)$ and $h^{(i)}(\cdot)$, $i=1,\ldots,l$, are linear, i.e., $f(x_k,u_k) = A x_k + B u_k$ and $ h^{(i)}(x_k) = C^{(i)} x_k$, $i=1,\ldots,l$, where $A$, $B$, $C^{(i)}$ are constant matrices of suitable dimensions.%
\end{enumerate} \vspace{.15cm}
\begin{enumerate}[\bf {A}3]
\item The PDFs of the measurement noises  are log-concave.
\end{enumerate} \vspace{.15cm}
Concerning assumption A3, it is worth noting that the class of probability distributions having  a log-concave PDF is quite general and includes, among others,
normal, exponential, uniform, logistic, and Laplace distributions.
Then, the following result, whose proof is in the Appendix, can be stated.
\vspace{.15cm}
\begin{proposition}
If assumption A2 holds, then the set $\mathbb X_k$ is an open convex polyhedron for any $k$. If in addition also assumptions A1 and A3 hold, then
the cost function (\ref{MH}) with arrival cost (\ref{eq:arrival}) is convex.
\mbox{   }\hfill $\square$
\end{proposition}
Notice that the convexity of the cost function~(\ref{MH}) is guaranteed also when assumption A1 is replaced by the milder requirement that the PDF of the process disturbance is log-concave.

\section{Finite element approximation and dynamic field estimation}

In this section, we consider the problem of reconstructing a two-dimensional dynamic field, sampled with a network of threshold sensors arbitrarily deployed over the spatial domain of interest $\Omega$. The process is governed by the following parabolic PDE:
\begin{equation}
\dfrac{\partial c}{\partial t} - \lambda \nabla^2 c  ~=~ 0 \,\,\, \mbox{in} \,\, \Omega
\label{PDE}
\end{equation}
which models various physical phenomena, such as the spread of a pollutant in a fluid.
In (\ref{PDE}):
$c(\xi,\eta,t)$ represents the space-time dependent substance concentration;
$\lambda$ is the constant diffusivity of the medium;
$\nabla^2 = {\partial^2 } / {\partial \xi^2} + {\partial^2 } / {\partial \eta^2}$ is the Laplace operator; $(\xi,\eta) \in \Omega$  and $t \in \mathbb{R}$ denote the planar Cartesian coordinates and, respectively, the continuous time instant.
Furthermore, let us assume mixed boundary conditions, i.e. (i) a non-homogeneous Dirichlet condition
\begin{equation}
c = \psi \,\, \mbox{on} \,\, \partial \Omega_D,
\label{Dbc}
\end{equation}
which specifies a constant-in-time value of concentration on the boundary $\partial \Omega_D$, and (ii) a homogeneous Neumann condition on $\partial \Omega_N = \partial \Omega \setminus \partial \Omega_D$, assumed impermeable to the contaminant, so that
\begin{equation}
\frac{\partial c}{\partial \upsilon} = 0 \,\, \mbox{on} \,\, \partial \Omega_N,
\label{Nbc}
\end{equation}
where $\upsilon$ is the outward pointing unit normal vector of
$\partial \Omega_N$.

The objective is to estimate the dynamic field of interest $c(\xi,\eta,t)$, given pointwise-in-time-and-space threshold measurements of the field itself.
The PDE system (\ref{PDE})-(\ref{Nbc}) is spatially discretized with a mesh of finite elements over $\Omega$ via the Finite Element (FE) approximation described in previous work on dynamic field estimation. \cite{source,ECC15,TAC17} Specifically, the domain $\Omega$ is subdivided into a suitable set of non overlapping regions, or elements, and a suitable set of basis functions $\phi_{j} (\xi , \eta)$,  $j=1,\ldots,n_\phi$, is defined on such elements. The choices of the basis functions $\phi_{j}$ and of the elements are key points of the FE method. In the specific case under investigation, the elements are triangles in 2D and define a FE mesh with vertices $({\xi}_j, \eta_j) \in \Omega, j=1,\ldots,n_\phi $. Then each basis function $\phi_{j}$ is a piece-wise affine function which vanishes outside the FEs around $({\xi}_j, \eta_j)$ and such that $\phi_{j}({\xi}_i, \eta_i)=\delta_{ij}$, $\delta_{ij}$ denoting the Kronecker delta. In order to account for the mixed boundary conditions, the basis functions are supposed to be ordered so that
the first $n$ correspond to vertices of the mesh which lie either in the interior of $\Omega$ or on $\partial \Omega_N$, while the last $n_\phi-n$ correspond
to the vertices lying on $\partial \Omega_D$. Accordingly, the unknown function $c(\xi, \eta,t)$
is approximated as
\begin{equation}
c(\xi,\eta,t) \approx \sum_{j=1}^{n} \phi_{j}(\xi, \eta) \, c_j(t) + \sum_{j=n+1}^{n_\phi} \phi_{j}(\xi, \eta) \, \psi_j
\label{EXPA}
\end{equation}
where $c_j(t)$ is the unknown expansion coefficient of the function $c(\xi, \eta,t)$ with respect to time $t$ and basis function $\phi_j(\xi,\eta)$, and $\psi_j$ is the known expansion coefficient of the function $\psi(\xi, \eta)$ only relative to the basis function $\phi_j(\xi,\eta)$. Notice that the second summation in
(\ref{EXPA}) is needed so as to impose the non-homogeneous Dirichlet condition (\ref{Dbc}) on the boundary $\partial \Omega_D$. The PDE (\ref{PDE}) can be recast into the
following integral form:
\begin{equation}
\int_\Omega \frac{\partial c}{\partial t} \varphi \, d\xi d\eta  \, - \,
\lambda \int_\Omega ~\nabla^2 c~ \varphi \, d\xi d\eta  =0
\end{equation}
where $\varphi(\xi,\eta)$ is a generic space-dependent weight function. By applying the Green's identity, one obtains:
\begin{equation}\label{eq:Green}
\int_\Omega \frac{\partial c}{\partial t} \varphi \, d\xi d\eta + \lambda
\int_\Omega \nabla^T c ~\nabla \varphi \, d\xi d\eta
- \lambda
\int_{\partial \Omega} \frac{\partial c}{\partial \upsilon} \varphi \,
d\xi d\eta
= 0 \, .
\end{equation}
By choosing the test function $\varphi$ belonging to the selected basis functions and exploiting the approximation (\ref{EXPA}), the Galerkin weighted residual method is then applied and the following equation is obtained
\begin{equation}\label{eq:Galerkin}
\sum_{i=1}^{n}  \int_\Omega \phi_i \phi_j \, d\xi d\eta \, \dot{c}_i(t)  + \lambda \sum_{i=1}^{n}  \int_\Omega \nabla^T \phi_i ~\nabla \phi_j \, d\xi d\eta \, c_i (t) + \lambda  \sum_{i=n+1}^{n_\phi} \int_\Omega \nabla^T \phi_i ~\nabla \phi_j \, d\xi d\eta \, \psi_i = 0
\end{equation}
for $j = 1, \ldots, n$. In the latter equation the boundary integral of equation (\ref{eq:Green}) is null 
thanks to the homogeneous Neumann condition (\ref{Nbc}) on $\partial \Omega_N$ and to the fact that, by construction, the basis functions $\phi_j$, $j = 1, \ldots, n$, vanish on $\partial \Omega_D$.
The interested reader is referred to the standard literature \cite{Brenner96} for further details on the FEM theory, in particular on how to convert the case of inhomogeneous boundary conditions to the homogeneous one.

By defining the state vector $x \triangleq {\rm col} (c_1 , \ldots , c_n) $ and the vector of boundary conditions $\gamma \triangleq {\rm col} (\psi_{n+1}, \ldots, \psi_{n_\phi})$, equation (\ref{eq:Galerkin}) can be written in the more compact form
\begin{equation}
M \dot x (t) + S x(t) + S_D \gamma = 0
\end{equation}
where $S$ is the so-called stiffness matrix, $M$ is the mass matrix,  and $S_D$ captures the physical interconnections among the vertices affected by boundary condition (\ref{Dbc}) and the remaining nodes of the mesh. By applying, for example, the implicit Euler method, the latter equation can be discretized in time, thus obtaining the
linear discrete-time model
\begin{equation}
x_{k+1} = A \, x_k + B \, u + w_k\,,
\label{dt-sys}
\end{equation}
where
\begin{eqnarray}
A &=& \left[ I + T_s ~M^{-1}S\right]^{-1} \\
B &=& \left[ I + T_s ~M^{-1}S\right]^{-1}M^{-1} T_s \\
u &=& - S_D ~\gamma ~.
\label{eq:def}
\end{eqnarray}
$T_s$ is the sampling interval, and $w_k$ is the process disturbance taking into account also the space-time discretization errors. Notice that the linear system (\ref{dt-sys}) has dimension $n$ equal to the number of vertices of the mesh not lying on $\partial \Omega_D$. The system (\ref{dt-sys}) is assumed to be monitored by a network of $l$ threshold sensors. Each sensor, before threshold quantization is applied, directly measures the pointwise-in-time-and-space concentration of the contaminant in a point $(\xi_i,\eta_i)$ of the spatial domain $\Omega$. By exploiting (\ref{EXPA}), such a concentration can be written as a linear combination of the concentrations on the grid points in that
\begin{equation}
c(\xi_i,\eta_i, kT_s) \approx C^{(i)} x_k + D^{(i)} \gamma \,,
\end{equation}
where
\begin{eqnarray}
C^{(i)} &=& \left [ \phi_1(\xi_i,\eta_i) \; \cdots \; \phi_n(\xi_i,\eta_i) \right ] \\
D^{(i)} &=& \left [ \phi_{n+1}(\xi_i,\eta_i) \; \cdots \; \phi_{n_\phi}(\xi_i,\eta_i) \right ].
\end{eqnarray}
Hence the resulting output function takes the form
\begin{equation}
z_{k}^{(i)} = c(\xi_i,\eta_i,kT_s)   - D^{(i)} \gamma + v_{k}^{i} = C^{(i)} x_{k} + v_{k}^{i},\hspace{3mm}i=1,\ldots,l \label{z}
\end{equation}
where the constant $D^{(i)} \gamma$ is subsumed into the threshold $\tau^{(i)}$, so that assumption A2 is fulfilled.

\section{Fast MH-MAP filter for field estimation of large-scale systems}

In order to achieve a good approximation of the original continuous field, a large number of basis functions need to be used in the expansion (\ref{EXPA}). Hence, in general the FEM-based space discretization gives rise to a large-scale system possibly characterized by thousands of state variables, equal to the number of the vertices of the mesh not lying on the boundary $\partial\Omega$. This means that a direct application of the MH-MAP filter of Section 4 to system (\ref{dt-sys}) involves the solution, at each time instant, of a large-scale (albeit convex) optimization problem. Although today commercial optimization software can solve general convex programs of some thousands variables, the problem becomes intractable from a computational point of view when the number of variables (that is, the number of vertices of the FE grid) is too large. Further, even when a solution to the large-scale optimization problem can be found, the time required for finding it may not be compatible with real-time operations (recall that the MH-MAP filter requires that each optimization terminates within one sampling interval).

In this section, we propose a more computationally efficient and faster version of the MH-MAP filter for the real-time estimation of a dynamic field that is based on the idea of decomposing the original large-scale problem into simpler, small-scale, subproblems by means of a two-stage estimation procedure.
The proposed method allows one to efficiently solve the problem of estimating the state (ideally infinite-dimensional) of a spatially-distributed dynamical system just by using sensors with minimal information content, such as threshold sensors. The fast version of the aforementioned MH-MAP filter, which can be suitable for large-scale systems, will split the estimation problem into two main steps.
\begin{enumerate}[(1)]
  \item Estimation of the local concentration in correspondence of each threshold sensor by means of $l$ independent MH-MAP filters. The concentration estimates provided by each local MH-MAP filter allows one to recast the threshold (binary-valued) measurements as \textit{linear} real-valued pseudo-measurements.
  \item Field estimation over a mesh of finite elements defined over the (spatial) domain $\Omega$ on the basis
  of the linear pseudo-measurements provided by the local filters in step\,1. For this purpose, any linear filtering technique suitable for large-scale systems can be used (see e.g.\,the finite-element Kalman filter  \cite{TAC17}). In this paper, field estimation is performed by minimizing a single quadratic MH cost function for linear systems.
\end{enumerate}
This solution turns out to be more computationally efficient as compared to a direct application of the MH-MAP filter to system (\ref{dt-sys}) in that: (i) the number of threshold sensors spread over the domain $\Omega$ is typically much smaller than the number of vertices of the FE mesh (i.e. $l\ll n$); (ii) as it will be clarified in the
following, each local MH-MAP filter in step 1 involves the solution of a convex optimization problem with a greatly reduced number of variables.

\subsection{Step~1}

Let us examine in more detail step~1 of the fast MH-MAP filter. To this end,
let us denote by $\sigma_{k}^{(i)}$ the value of the concentration in correspondence of sensor $i$
at the $k$-th sampling instant, i.e. $\sigma_{k}^{(i)} = c(\xi_i,\eta_i, k T_s)$, and let
$\sigma^{(i)}_{m,k}$ denote the value of the $m$-th time-derivative of such a concentration, so that
\begin{equation}\label{eq:taylor}
\sigma^{(i)}_{m,k} = \left . \frac{\partial^m}{\partial t^m} c(\xi_i,\eta_i,t) \right |_{t = k T_s}.
\end{equation}
Under the hypothesis of a small enough sampling interval, the dynamical evolution of the propagating field in correspondence of each binary sensor can be approximated by resorting to a truncated Taylor series expansion, i.e.
\begin{equation}
\sigma_{k+1}^{(i)} \approx \sigma_{k}^{(i)} + \sum_{m=1}^M \frac{T_s^m}{m!} \sigma^{(i)}_{m,k}
\end{equation}
Then, the local dynamics of the concentration in correspondence of the sensor $i$ can be described by a linear dynamical system with state $\chi_{k}^{(i)} \triangleq {\rm col} \left( \sigma_{k}^{(i)},\sigma_{1,k}^{(i)} , \ldots, \sigma_{M,k}^{(i)}\right)$ and state equation
\begin{equation}\label{eq:chi}
\chi_{k+1}^{(i)} = \tilde A \, \chi_{k}^{(i)} + w_k^{(i)}\,,
\end{equation}
where the matrix $\tilde A$  is obtained by time-discretization of (\ref{eq:taylor}) with sampling interval $T_s$ and $w_k^{(i)}$ is the disturbance acting on the local dynamics with zero mean and inverse covariance $\tilde G$. Notice that models like (\ref{eq:chi}) are widely used in the construction of filters for the estimation of time-varying quantities whose dynamics is unknown or too complex to be modeled (for instance, they are typically used in tracking as motion models of moving objects~\cite{Bar-Shalom}). With this respect, a crucial assumption for the applicability of this kind of models is that the sampling interval is sufficiently small as compared to the time constants characterizing the variation of the quantities to be estimated. Hence, their application in the present context is justified by the fact that, in practice, (binary) concentration measurements can be taken at a high rate so that between two consecutive measurements only small variations can occur.

In model (\ref{eq:chi}), the simplest choice amounts to take $M=0$ and $w^{(i)}_k$ as a Gaussian white noise, which corresponds to approximate the concentration as nearly constant (notice that, in this case, we have $\tilde A = 1$). Instead, by taking $M=1$, we obtain a nearly-constant derivative model with state transition matrix
\begin{equation}
\tilde A = \left [ \begin{array}{cc} 1 & T_s \\ 0 & 1 \end{array} \right],
\end{equation}
which is equivalent to the nearly-constant velocity model widely adopted for moving object tracking. \cite{Bar-Shalom} Clearly, each local model (\ref{eq:chi}) is related to the $i$-th binary measurement via the measurement equation
\begin{eqnarray}
z_{k}^{(i)} &=& \widetilde C \, \chi_{k}^{(i)} + v_{k}^{(i)} \\
y_{k}^{(i)} &=& g^{(i)} \left ( z_{k}^{(i)} \right )\,, \label{meas:chi}
\end{eqnarray}
where
\begin{equation}
\widetilde C = [1 \; 0 \; \cdots \; 0] \, .
\end{equation}
Then, for each sensor $i$ and time instant $k$, the minimization of the following MH-MAP cost function is addressed
\begin{eqnarray}\label{eq:cost_step1}
\widetilde J_{k}^{(i)} (\chi_{k-N:k}^{(i)}) &=& \|\chi_{k-N}^{(i)} - \overline{\chi}_{k-N}^{(i)}\|^{2}_{\widetilde \Psi} + \displaystyle{\sum_{j=k-N}^{k}}\|\chi_{j+1}^{(i)} - \tilde A \, \chi_{j}^{(i)} \|^{2}_{\widetilde G}\nonumber \\
&-&\displaystyle{\sum_{j=k-N}^{k}}\left\{ (1-y_{j}^{(i)}) \ln F^{(i)} \left (\tau^{(i)}- \widetilde C \,  \chi_{j}^{(i)}  \right )+
y_{j}^{(i)} \ln \left[ 1 - F^{(i)} \left (\tau^{(i)}- \widetilde C \,  \chi_{j}^{(i)}  \right ) \right] \right\},
\end{eqnarray}
where $\chi_{k-N:k}^{(i)} \triangleq {\rm col}\left( \chi_{k-N}^{(i)} , \ldots, \chi_{k}^{(i)}\right)$
and $\overline{\chi}_{k-N}^{(i)}$ is the estimate of the local state at time  $k-N$ computed at the previous iteration.

In conclusion, at any time instant $k = N, N+1, \ldots$, for any threshold sensor $i$ the following problem has to be solved. \\ \\
\textbf{Problem $\widetilde E_k^{(i)}$:} Given the prediction $\overline{\chi}_{k-N}^{(i)}$ and the measurement sequence $\{ y^{(i)}_{k-N}, \ldots, y^{(i)}_k \}$, find  the optimal estimates $\hat{\chi}_{k-N|k}^{(i)},\ldots,\hat{\chi}_{k|k}^{(i)}$ that minimize the cost function $\widetilde J_{k}^{(i)}(\chi_{k-N:k}^{(i)})$ in (\ref{eq:cost_step1}). \\ \\
As before, the propagation of the estimation problem from time $k-1$ to time $k$ is ensured by choosing $\overline{\chi}_{k-N}^{(i)} = \hat{\chi}_{k-N|k-1}^{(i)}$.
The number of variables involved in each of such optimization problems is $(M+1)(N+1)$ and, in view of (\ref{eq:chi}) and (\ref{meas:chi}), the cost function $\widetilde J_{k}^{(i)}$ is \textit{convex} according to Proposition~1. Hence, basically, step 1 amounts to solving $l$ convex optimization problems of low/moderate size.

\subsection{Step~2}

In step 2, the concentration estimates $\hat{\sigma}_{j|k}^{(i)} = \widetilde{C} \hat{\chi}_{j|k}^{(i)}$, $j= k-N, \ldots, k$, obtained by solving Problem $\widetilde E^{(i)}_k$ in correspondence of any threshold sensor $i$, are used as linear pseudo-measurements in order to estimate the whole concentration field over the spatial domain $\Omega$. By resorting again to the
FE approximation of Section V, the vector of coefficients $x_k$ can be estimated, for example, by minimizing a quadratic MH
cost function of the form:
\begin{equation}\label{eq:cost_function_2}
\overline J_{k}(X_{k-N:k}) = \|x_{k-N} - \overline{x}_{k-N}\|^{2}_{\Psi} + \sum_{j=k-N}^{k}\|x_{j+1}-Ax_{j}-Bu\|^{2}_{G}
+\sum_{j=k-N}^{k}\sum_{i=1}^{l}\|\hat{\sigma}_{j|k}^{(i)}-C^{(i)}x_{j} - D^{(i)} \gamma \|^{2}_{\Xi^{(i)}}
\end{equation}
where the quantities $A$, $B$, $\gamma$, $C^{(i)}$, $D^{(i)}$, for $i=1,\ldots,l$, are obtained by means of the
FE method as in Section V. Notice that each term weighted by the positive scalar $\Xi^{(i)}$ penalizes the distance of the concentration $C^{(i)} x_{j} + D^{(i)}\gamma$, estimated through the FE approximation from the concentration
$\hat{\sigma}_{j|k}^{(i)}$ at step~1 on the basis of the threshold measurements.
In practice, each weight $\Xi^{(i)}$ can be set approximately equal to the inverse of the variance of the estimation error in the step~1 of the procedure, computed for example by means of numerical simulations.
Once these weights have been fixed, the weight matrix $\Psi$ in the arrival cost can be tuned so as to ensure the stability of the estimation error. We refer to the related literature on MHE \cite{receding,NLMHE,CDC} for a discussion on this issue.
The prediction $\overline{x}_{k-N}$ is computed in a recursive way as shown in the previous sections. Accordingly, at any time instant $k = N, N+1, \ldots$, after the application of step~1 the following problem has to be addressed. \\ \\
\textbf{Problem $\overline E_k$:} Given the prediction $\overline{x}_{k-N}$ and the optimal estimates
$\{\hat{\sigma}_{k-N|k}^{(i)},\ldots,\hat{\sigma}_{k|k}^{(i)}\}$, $i = 1, \ldots, l$, obtained by solving Problem $\widetilde E_k^{(i)}$, find the optimal estimates $\hat{x}_{k-N|k},\ldots,\hat{x}_{k|k}$ that minimize the cost function $\overline J_{k}(X_{k-N:k})$ in (\ref{eq:cost_function_2}). \\ \\
Since the cost function (\ref{eq:cost_function_2}) depends quadratically on the states $\{x_{k-N},\ldots,x_{k}\}$, the above estimation problem admits a closed-form solution at any time $k$.
Hence, the computational effort needed to perform step 2 of the fast MH-MAP filter turns out to be limited.
Recall also that Step 1 of the proposed approach involves the solution of $l$ nonlinear nonquadratic convex optimization problems in $(K+1)(N+1)$ variables where $l$ is the number of threshold sensors, $N$ is the moving-horizon length and $K+1$ the low-order of the assumed local dynamics.
This means that the overall algorithm is computationally efficient if compared to a direct application of the MH-MAP filter on the large-scale system arising from the FE discretization,
which would require the solution of a large-scale optimization problem with a number of variables equal to $(N+1)n$ where $n$ is the order of the adopted FE model. Such improvement becomes even more relevant if  compared with other common solutions already present in the literature. As a matter of fact, it is worth noting that approaches relying on particle filtering \cite{Djuric_1,Djuric_2} cannot be applied to solve estimation problems relative to large-scale systems, and deterministic approaches \cite{Irr-sampling,Gherardini2,Wim2006,Bai} require the optimization of a piece-wise cost function, due to the highly discontinuous nature of the outcomes from threshold sensors. Although piece-wise optimization can be solved by means of sequential quadratic programming, the corresponding computational cost is feasible only if the number of state variables is not too large.

\section{Numerical results}

Simulation experiments concerning state estimation of a spatially distributed system
are used to demonstrate the effectiveness of the proposed (both standard and computationally fast) MH-MAP approach.
The process under consideration is modeled by a diffusion PDE described by \eqref{PDE} with known constant diffusivity $\lambda = 5 \times 10^{-8} ~ [m^2/s]$
and initial condition $x_{0} = 0_n ~ [g/m^2]$. Diffusion is a fundamental transport process in fluid dynamics which governs
the movement of a substance from a region of high concentration to a region of low concentration.
The discrete-time state space system (\ref{dt-sys})-(\ref{z}) is obtained by first using FEM for
spatial discretization of the PDE with a mesh of $1695$ triangular elements ($915$ vertices) generated by the Matlab PDE Toolbox, and then via time-discretization of the resulting model with fixed sampling interval $T_s = 1 ~ [s]$.

\begin{figure}[h!]
	\centering
	\includegraphics[scale=0.325]{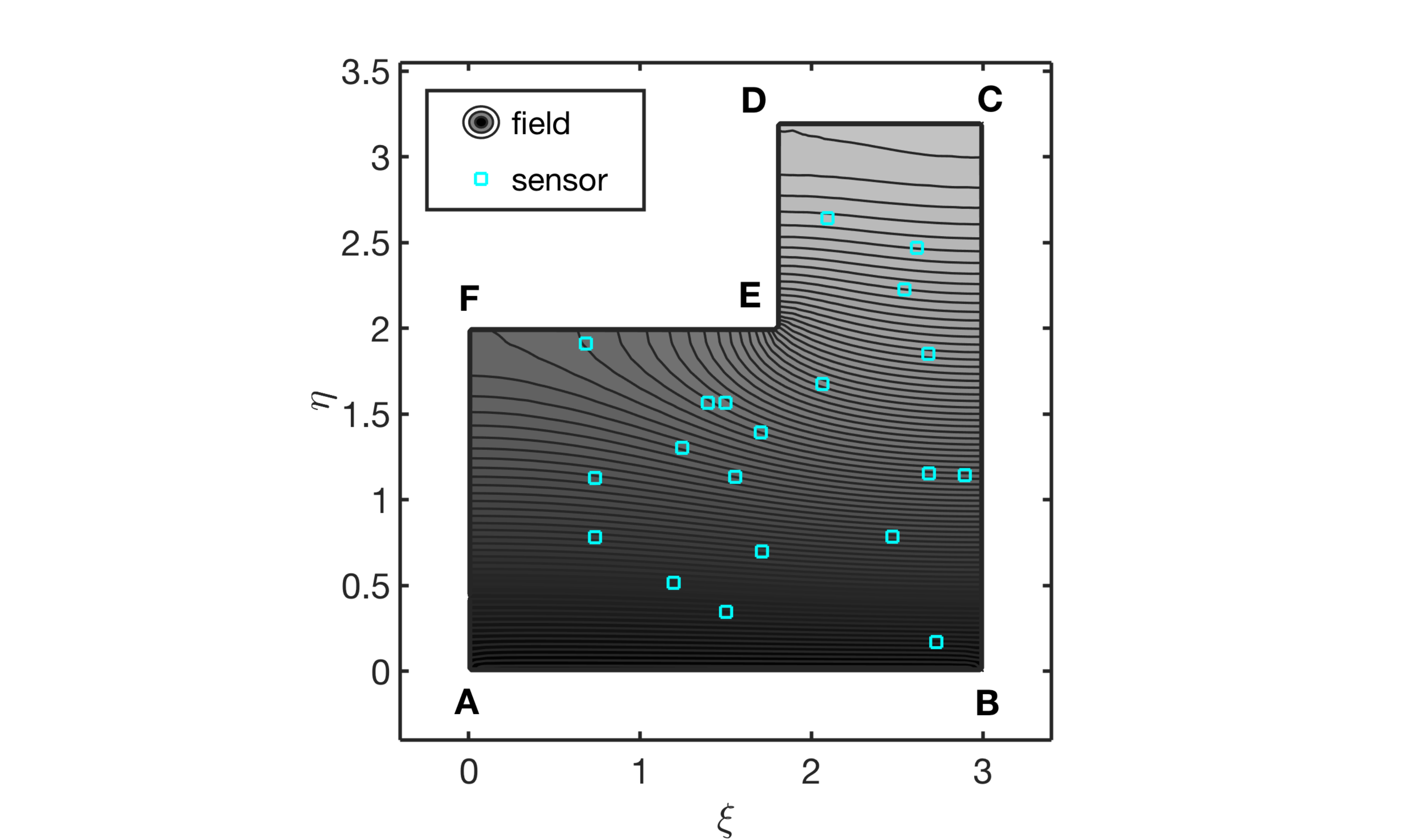}
	\caption{Diffusive field monitored by a constellation of 20 threshold sensors (cyan $\square$) randomly deployed over the 2D bounded domain $\Omega$.}
	\label{fig:field200}
\end{figure}
\begin{figure}[h!]
	\vspace{-.2cm}
	\centering
	\includegraphics[scale=0.325]{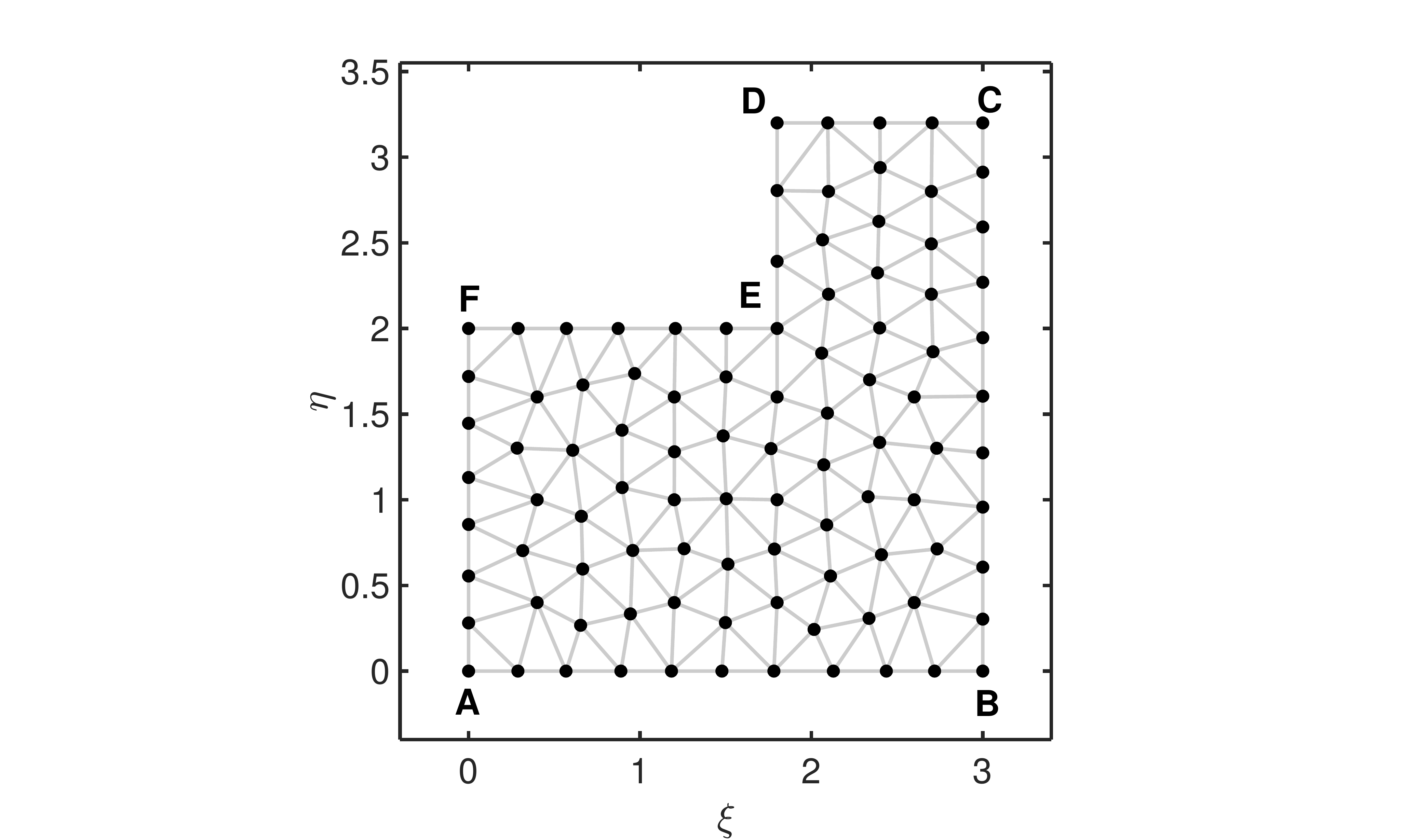}
	\caption{Triangular finite-element mesh (152 elements, 97 nodes) used by the proposed MH-MAP field estimators.}
	\label{fig:mesh}
\end{figure}

The \textit{true} dynamic field to be estimated is defined over an L-shaped spatial domain $\Omega$ covering an area of $7.44 ~ [m^2]$ (see Fig.\,\ref{fig:field200}). The problem is characterized by mixed boundary conditions, i.e. a condition of type (\ref{Dbc}) with $\gamma = 30 ~ [g/m^2]$ on the bottom end A--B, and no-flux condition (\ref{Nbc}) on the remaining five edges of boundary $\partial\Omega$. The \textit{true} concentrations are generated by recursively using (\ref{dt-sys}) on the simulated system. Then, threshold measurements are obtained by first corrupting the state variables with a Gaussian noise with variance $r^{(i)}$, and finally by applying a threshold $\tau^{(i)}$, different for each sensor $i$ of the network. Note that, in order to collect threshold measurements which are as informative as possible, $\tau^{(i)}, ~ i = 1,...,l$, are generated as uniformly distributed random numbers in the interval $[0.05,29.95]$, where $[0,30]$ is the range of nominal concentration values throughout each experiment. The duration of each simulation experiment is fixed to $1200 ~ [s]$ (120 samples).
In order to account for model uncertainty, the MH-MAP estimator is taken as one order of magnitude coarser and slower than the corresponding \textit{ground truth} simulator, by implementing a triangular mesh of ${n_\phi} = 97$ vertices (of which $n=89$ are internal) and $152$ elements (see Fig.\,\ref{fig:mesh}), and by running at a sample rate of $0.1 ~ [Hz]$.
It is worth to point out that such differences in the FE mesh resolutions and sampling rates between the ground-truth simulator and MH-MAP estimator clearly induce model mismatches in the simulation experiments, thus allowing to investigate the robustness of the proposed approach.
The time evolution of the field to be estimated is represented through the $n$-dimensional linear discrete-time model \eqref{dt-sys}, which describes the dynamics of the concentration field in correspondence of the $n$ nodes lying on the interior of $\Omega$. As discussed in Section V, based on these pointwise-in-space estimates and thanks to the FEM, it is possible to approximately reconstruct the overall state of the infinite-dimensional diffusive process \eqref{PDE}.
In this regard, we have set the following parameters: initial guess of the estimated field randomly generated as normally distributed with mean $\overline{x}_{0} = 5 \cdot {1}_n ~ [g/m^2]$ and variance $P= 10 ~ I_n$; measurement noise variance $r^{(i)} = 0.1$;
moving window of size $N = 15$;
total number of threshold sensors $l=20$;
 $G = 10 ~ I_n$ in \eqref{prob};
 weight matrices $\widetilde{\Psi} = 10^3 ~ I_n$ and $\widetilde{Q} = 10^{2} ~ I_n$ in \eqref{eq:cost_step1} (where  $1_{n}$ indicates the $n-$dimensional vectors with all unit entries, while $I_n$ denotes the $n-$dimensional identity matrix). Additionally, in all the numerical experiments we assumed a nearly-constant concentration field, i.e. $\tilde A = 1$ in \eqref{eq:chi}.
Further, we have also assumed a mismatch on the measurement noise covariance by taking $r^{(i)} = 1$ in (\ref{eq7}) instead of the ground truth value $0.1$.

\begin{figure}[t]
	\centering
	\includegraphics[scale=0.275]{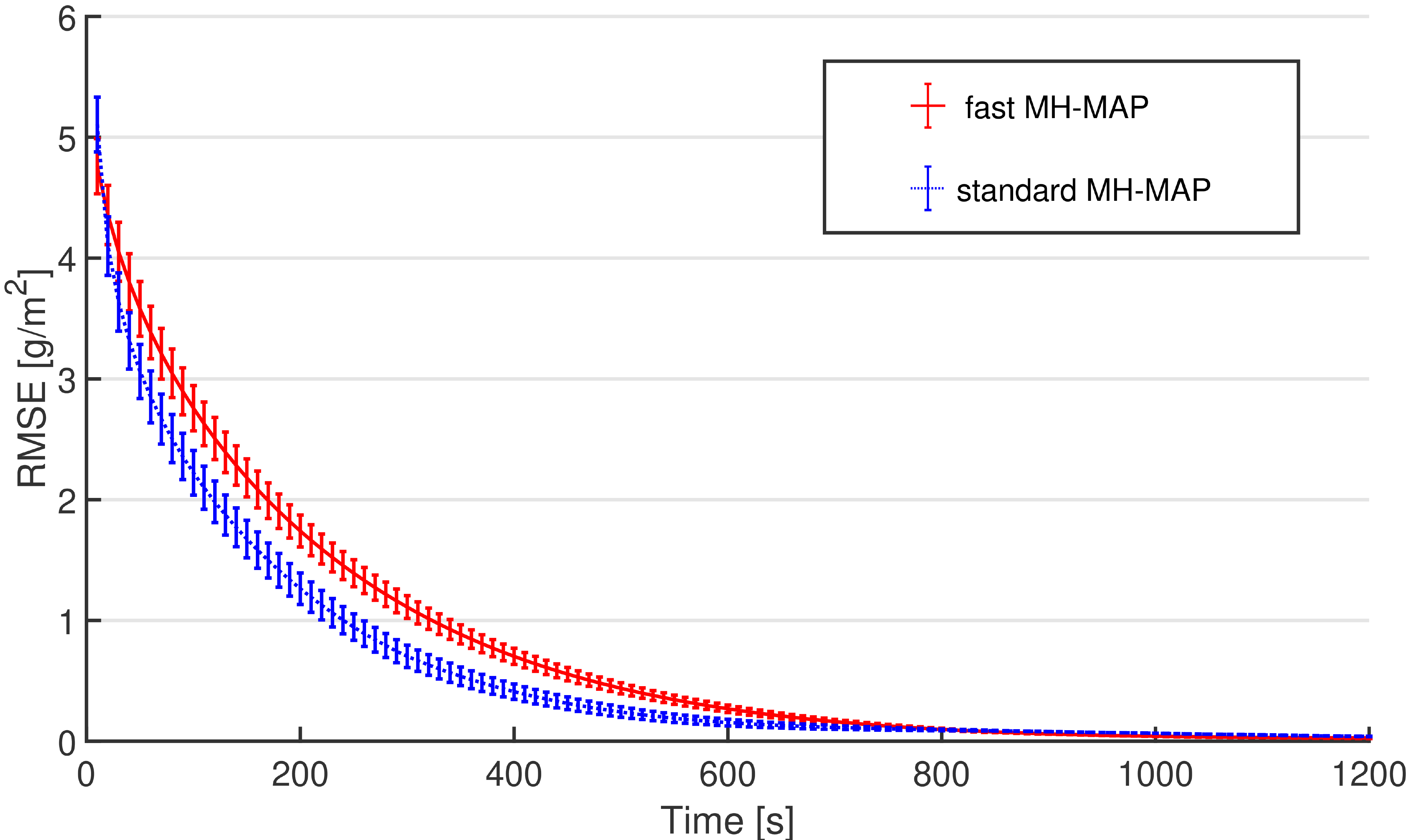}
	\caption{Comparison in terms of RMSE between the standard and fast MH-MAP state estimators as a function of time.}
	\label{fig:fastvsstand}
\end{figure}
\begin{figure}[t]
	\centering
	\includegraphics[scale=0.275]{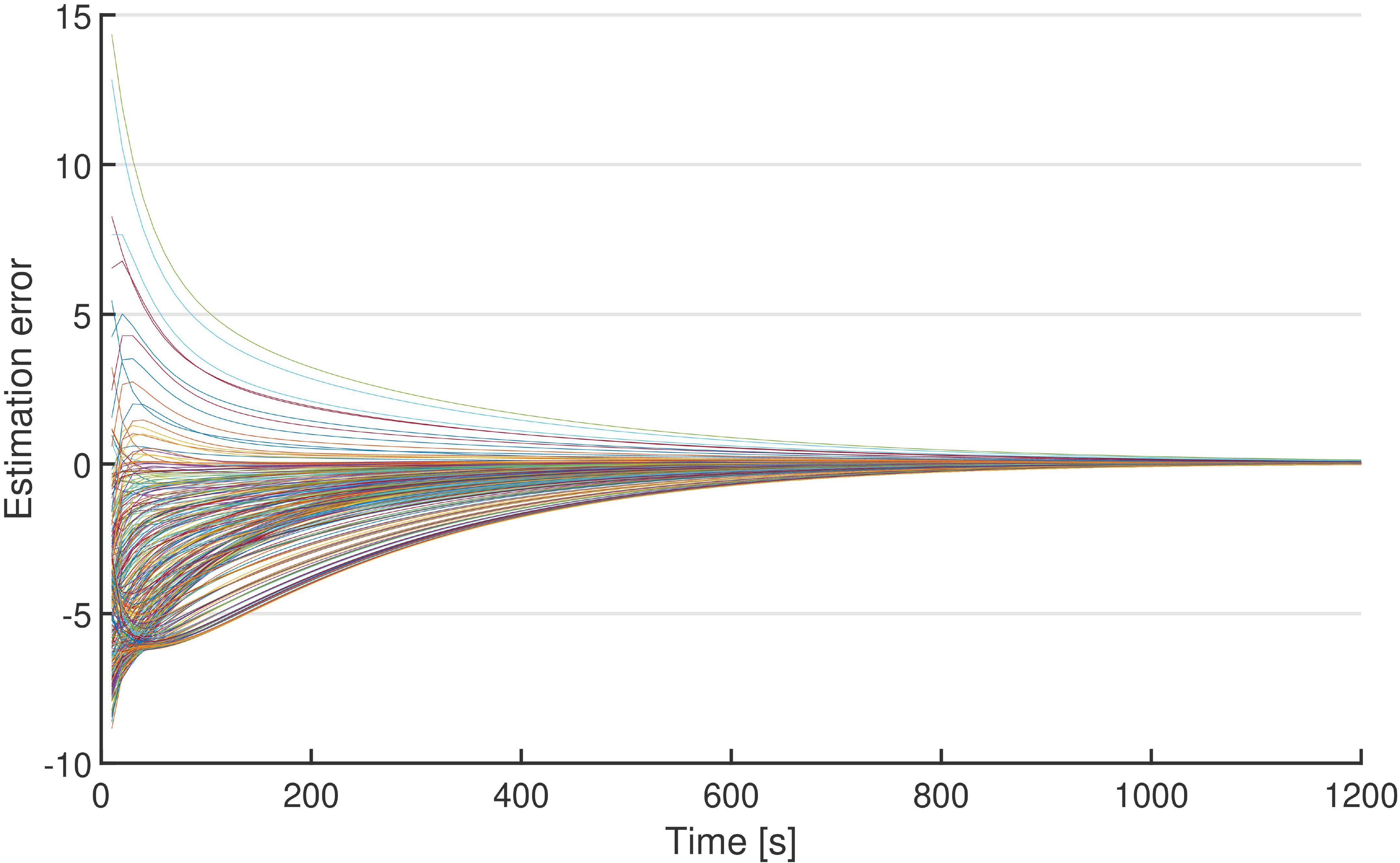}
	\caption{Fast MH-MAP filter: convergence of the concentration field estimates as simulation moves forward in time. The estimation error is computed for 304 sampling points evenly spread within the domain $\Omega$ in a single simulation run.}
	\label{fig:convergence}
\end{figure}
\begin{table}[t]
	\renewcommand{\arraystretch}{1.3}
	\caption{Computational performance: standard vs.\,fast MH-MAP filter}
	\centering
	\resizebox{9cm}{!}{
	\begin{tabular}{c||c||c}
		\hline\bfseries MH-MAP filter & \bfseries Optimization step [s]  & \bfseries Total CPU time [s]
		\\\hline\hline Standard & 21.7178 & 22.5529 \\
		\hline Fast & 0.4263 & 0.5391 \\
		\hline
	\end{tabular}
}
	\label{tab:1}
\end{table}

The performance of the proposed MH-MAP state estimators (both the standard and fast versions) is given in terms of \textit{Root Mean Square Error} (RMSE) of the estimated concentration field, i.e.
\begin{equation}\label{64}
\text{RMSE}(k)=\left(\sum_{\ell=1}^{L}\frac{\|e_{k,\ell}\|^{2}}{L}\right)^{\frac{1}{2}}
\end{equation}
where $\|e_{k,\ell}\|$ is the norm of the estimation error at time $k$ in the $\ell-$th simulation run (averaged over $L$ independent Monte Carlo realizations) and the estimation error is computed on the basis of the estimate $\hat{x}_{k-N|k}$. Fig.\,\ref{fig:fastvsstand} shows the comparison in performance between the standard and fast MH-MAP filters implemented in MATLAB, in terms of the RMSE, and relative standard deviation with respect to Monte Carlo runs, of the estimated diffusive field as given by (\ref{64}). To obtain the RMSEs plotted in Fig.\,\ref{fig:fastvsstand}, the estimation error $e_{k,\ell}$ at time $k$ in the $\ell-$th simulation run has been averaged over $304$ sampling points (evenly spread within $\Omega$) and $L=100$ Monte Carlo realizations.
As shown in Fig.\,\ref{fig:fastvsstand}, the estimation accuracy and relative statistical variability of the standard and fast MH-MAP filters is comparable, therefore the fast implementation is preferable in large-scale problems, usually arising from discretization of PDEs, where its computational efficiency is a strong advantage. In addition, the computational times associated to state estimation using the standard or fast filter are presented in Table~\ref{tab:1}, which highlights the effort required by the optimization step with respect to the total CPU time in the two cases. It is clearly shown that the on-line calculation times (tested on an Intel Xeon CPU @ $3.30$ GHz, $16$ GB RAM) are reduced dramatically when the fast MH-MAP filter is adopted. In particular, the computational burden can be reduced up to two orders of magnitude by running the fast two-step approach instead of directly applying the MH-MAP method. Notice also that around $96 \%$ and, respectively, $80 \%$ of the total CPU time is devoted to the optimization step, and that the considered field estimation problem can be reliably solved by the fast algorithm under the given sampling rate.
Fig.\,\ref{fig:convergence} shows how the diffusive field estimates obtained by the computationally fast MH-MAP filter tend to the true concentration values as time goes by.

\begin{figure}[t]
	\centering
	\includegraphics[scale=0.32]{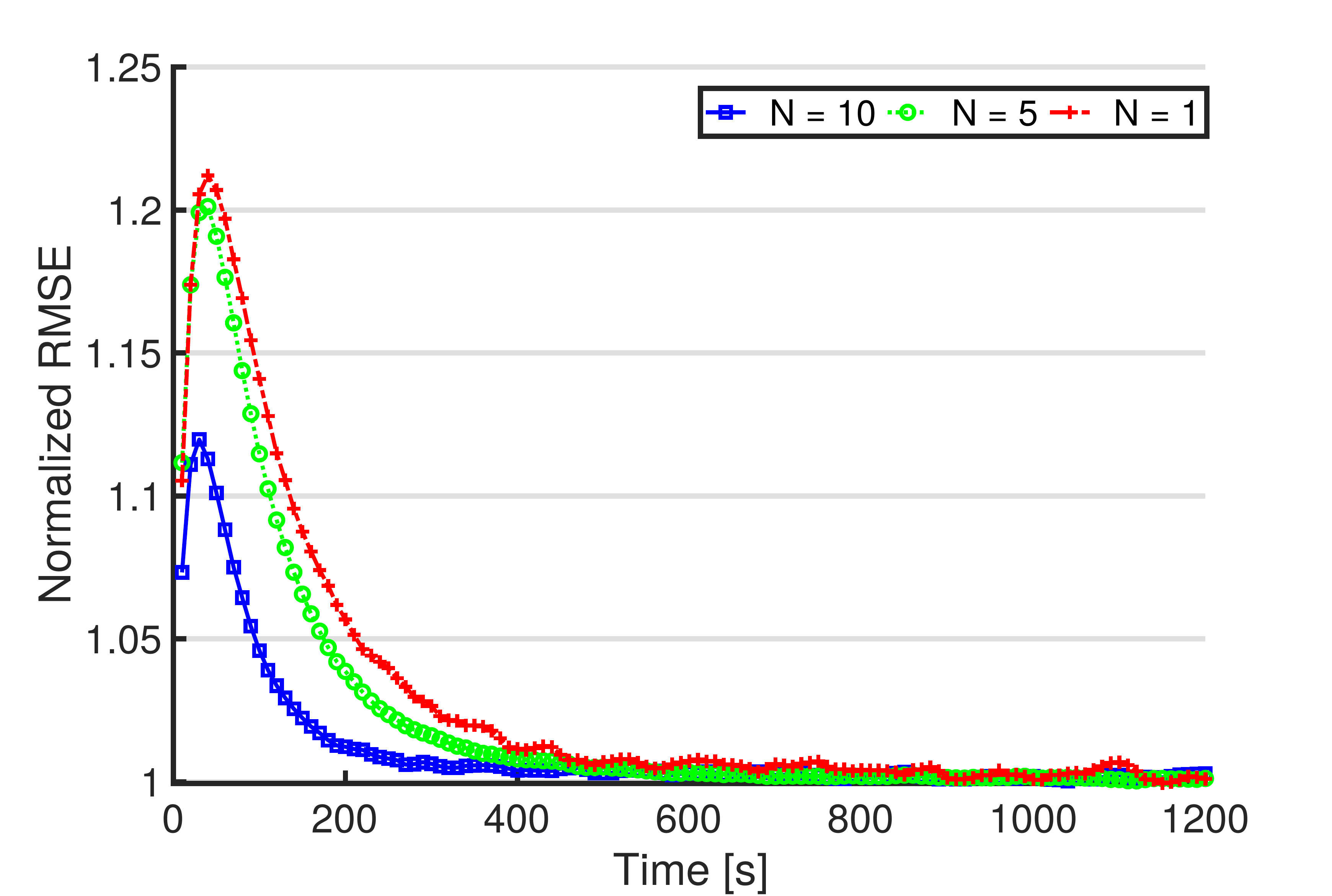}
	\caption{Increase of the RMSE of concentration estimates for varying  moving window size ($N=1,5,10$) compared to $N=15$.}
	\label{fig:normalizedH}
\end{figure}

\begin{figure}[t]
	\centering
	\includegraphics[scale=0.36]{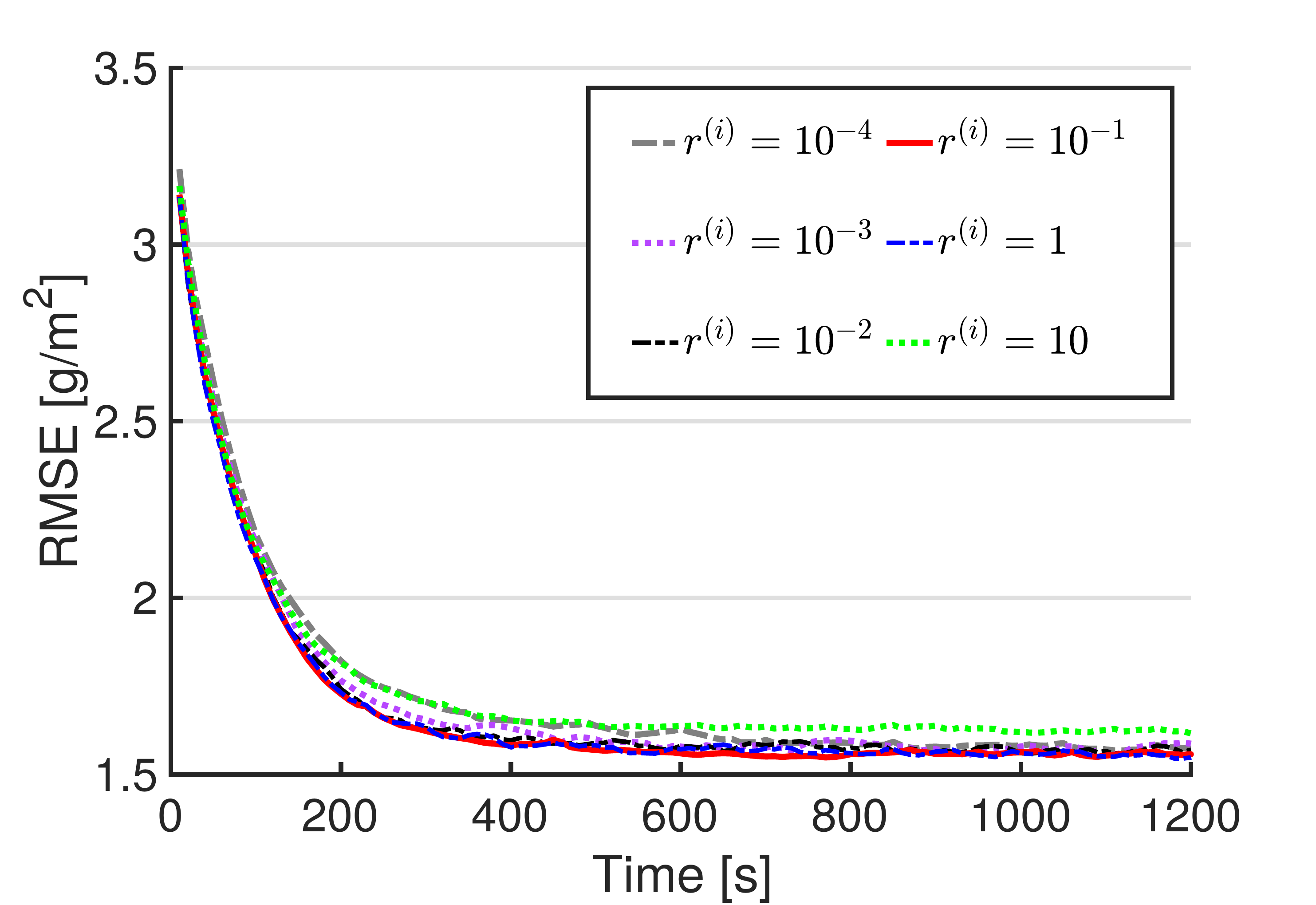}
	\caption{Comparison in terms of RMSE for varying measurement noise variance of the fast MH-MAP filter ($N=5$).}
	\label{fig:fig1}
\end{figure}

\begin{figure}[t]
	\centering
	\includegraphics[scale=0.35]{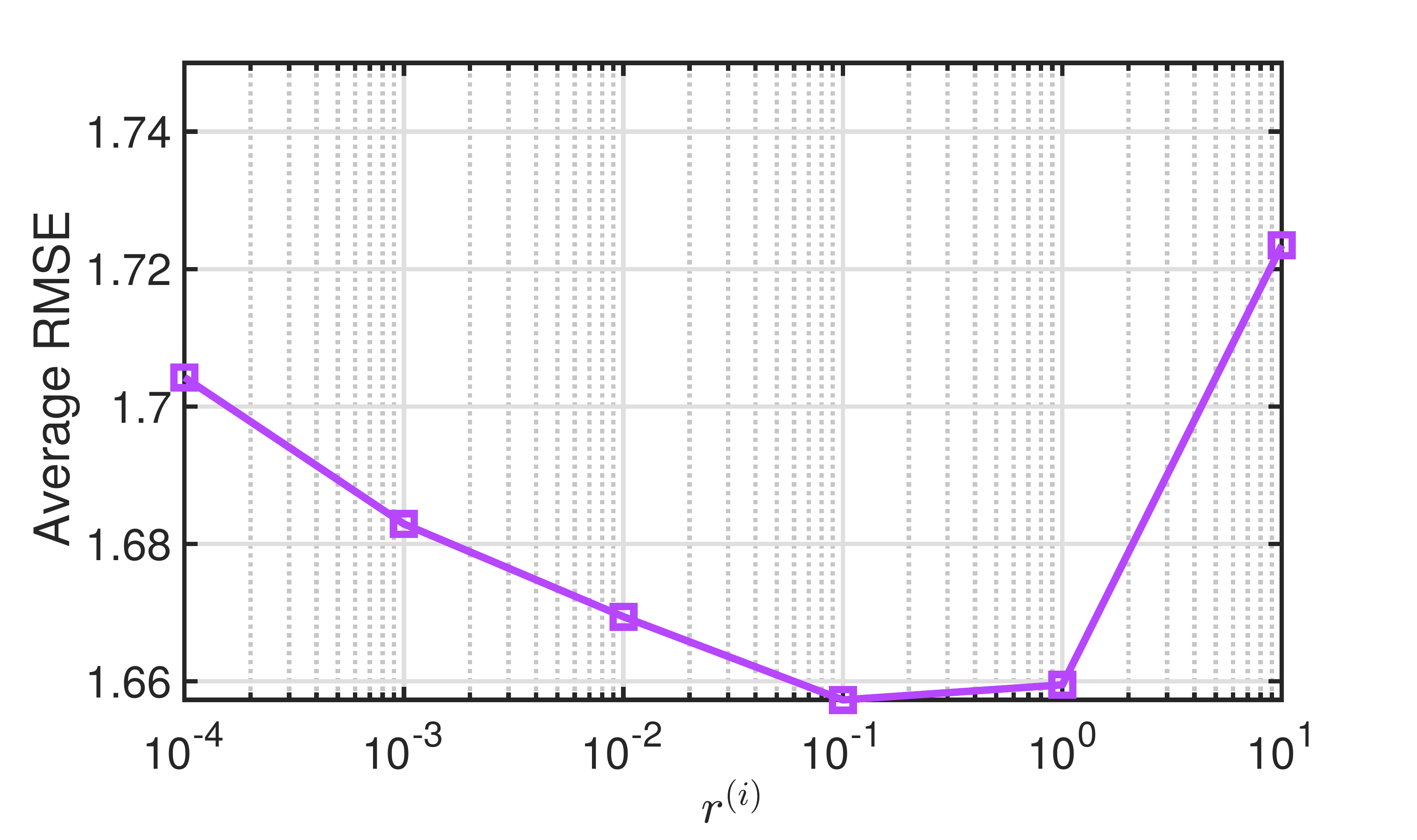}
	\caption{Fast MH-MAP filter: average RMSE over time as a function of the measurement noise variance, for a random network of 20 threshold sensors ($N=15$).
		As shown above, operating in a noisy environment turns out to be beneficial, for certain values of $r^{(i)}$, to the state estimation problem.}
	\label{fig:fig2}
\end{figure}

\begin{figure}[t]
	\centering
	\includegraphics[scale=0.225]{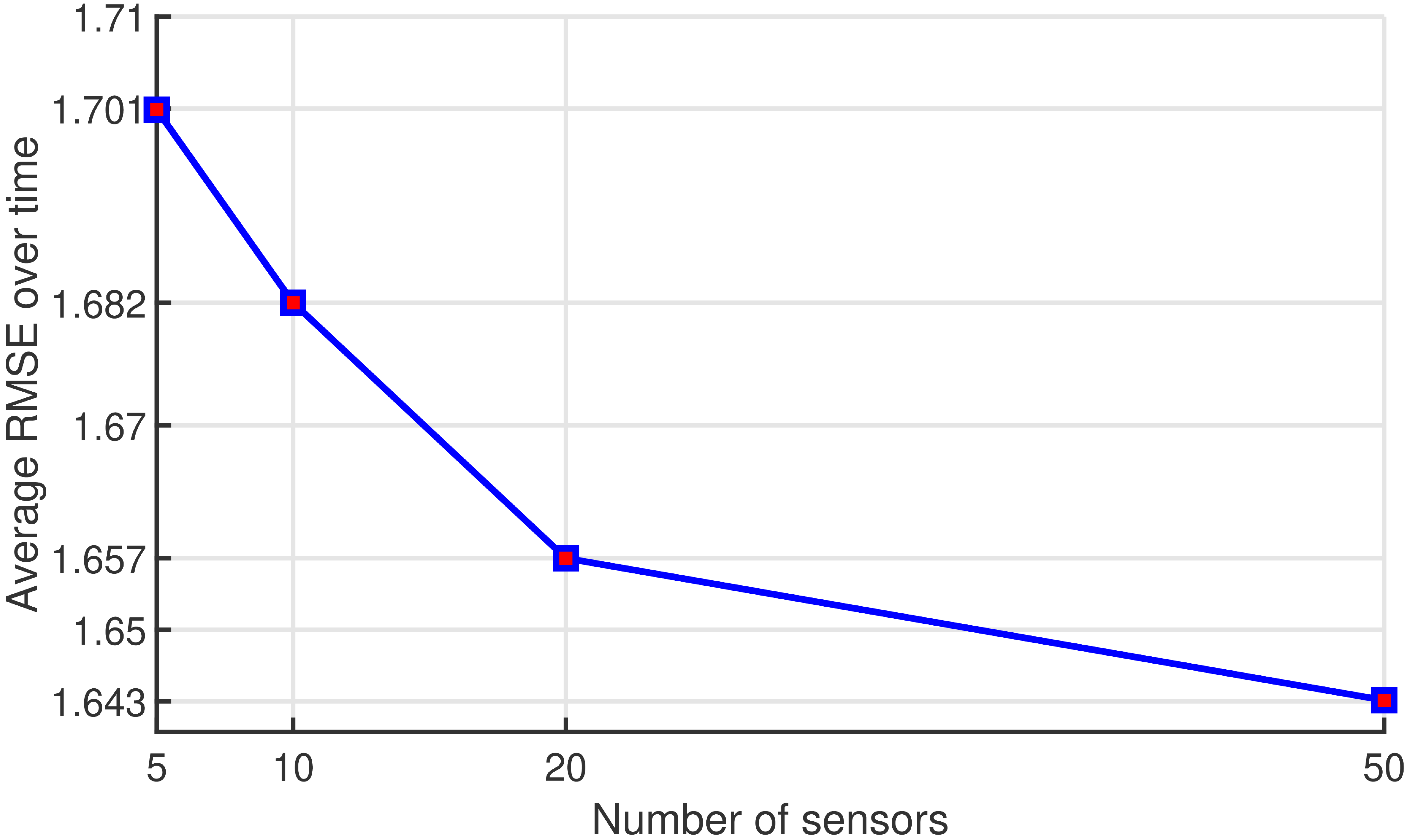}
	\caption{Fast MH-MAP filter: average RMSE over time as a function of the number of sensors deployed over the monitored area.}
	\label{fig:RMSE_nsens}
\end{figure}

Further simulations have been run in order to investigate the effect of the moving window length $N$, the measurement noise variance $r^{(i)}$, and the number of threshold sensors $l$ on the overall state estimation accuracy of the fast MH-MAP filter.
Its performance under different values of moving window length $N$ is illustrated in Fig.\,\ref{fig:normalizedH}. Specifically, Fig.\,\ref{fig:normalizedH} highlights the increase of RMSE that occurs when reducing the size of the moving window. It shows the RMSE for $N=1,5,10$ normalized over the estimation error obtained when $N=15$, i.e.\,the normalized RMSE (NRMSE), that is defined as
\begin{equation}
\text{NRMSE}_h=\frac{\text{RMSE}_{N=h}}{\text{RMSE}_{N=15}}, \quad \mbox{ for } h=1,5,10.
\end{equation}
Figs.\,\ref{fig:fig1}-\ref{fig:fig2} show the effect of measurement noise variance on the RMSE. Although the performance given varying values of $r^{(i)}$ is similar (see Fig.\,\ref{fig:fig1}), surprisingly the choice of an observation noise with higher variance can actually improve the overall quality of the field estimates. Such results are valid for both standard and fast MH-MAP filters (as shown in preliminary results \cite{ACC_binary} on standard MH-MAP field estimation), and they numerically demonstrate the validity of the above stated noise-assisted paradigm in problems of recursive state estimation using threshold measurements.
Finally, Fig. \ref{fig:RMSE_nsens} shows the evolution of the RMSE as a function of the number of threshold sensors available for field estimation.

\section{Conclusions}

State estimation with noisy threshold measurements has been addressed. The problem has been recast in a Bayesian framework in order to always ascribe to each threshold sensor a probabilistic information content encoded by the likelihood function. Accordingly, a MAP optimization approach has been undertaken by maximizing the conditional probability of the whole state trajectory given the whole sequence of threshold (binary) measurements. Since the dimension of such  MAP optimization clearly grows with time, an MH approximation over a moving time-window of suitable fixed length has been then adopted in order to make the problem solution computationally feasible. Further, it has been shown how the resulting MH-MAP state estimator can be efficiently applied to the monitoring of dynamic spatial fields by means of threshold sensors deployed over the area to be monitored.
Such fields are described by PDEs, which are spatially discretized with a mesh of finite elements over the spatial domain of interest by means of the FE method.
In order to efficiently estimate the time-space evolving field with a network of pointwise-in-time-and-space threshold sensors, we have proposed a faster version of the aforementioned MH-MAP filter, suitable for large-scale systems. The effectiveness of the proposed approach has been demonstrated via simulation experiments. The faster strategy has been shown to be able to provide similar levels of accuracy while requiring minimal computational effort compared to the standard MH-MAP filter, which is of great advantage in practical large-scale applications. Future work on the topic will possibly concern application of the MH-MAP state estimator  to target tracking with threshold proximity sensors.

\appendix

\section*{Proof of Proposition 1}

In order to show that $\mathbb X_k$ is an open convex polyhedron, let us consider the terms
 $- \ln F^{(i)}(\tau^{(i)}-h^{(i)} (x_{j}) ) = - \ln F^{(i)}(\tau^{(i)}-C^{(i)}x_{k})$  and $- \ln ( 1- F^{(i)}(\tau^{(i)}-h^{(i)} (x_{j})) ) =  - \ln ( 1 - F^{(i)}(\tau^{(i)}-C^{(i)}x_{j}) )$.
 Since the CDF $F^{(i)} (z)$ is non-negative and monotonically non-decreasing, the set of values $z$ such that $F^{(i)}(z) > 0$ is of the form $(a , \infty)$ for some $a \in \mathbb R \cup \{- \infty\}$.
 Hence, the domain of $- \ln F^{(i)}(\tau^{(i)}-C^{(i)}x_{k})$ has the form $\{ x_j \in \mathbb R^n : \tau^{(i)}-C^{(i)}x_{j} > a \}$ which defines an open hyperplane in $\mathbb R^n$.
 Similarly,  the set of values $z$ such that $1-F^{(i)}(z) > 0$ is of the form $(-\infty , b)$ for some $b \in \mathbb R \cup \{ \infty\}$.
 Hence, the domain of $- \ln ( 1 - F^{(i)}(\tau^{(i)}-C^{(i)}x_{j}) )$ has the form $\{ x_k \in \mathbb R^n : \tau^{(i)}-C^{(i)}x_{j} < b \}$ which also defines an open hyperplane in $\mathbb R^n$.
 Therefore, the set $\mathbb X_k$  is the intersection of a finite number of open hyperplanes which corresponds to an open convex polyhedron.

Concerning the convexity of the cost function, since the sum of convex functions is again convex, it is sufficient to show that each term of the cost is convex.
Clearly,  under assumptions A1 and A2,  each term $\| x_{j+1} - f(x_j, u_j) \|^2_G =  \| x_{j+1} - A x_j -B u_j \|^2_G$ is convex.
Consider now the terms $ - \ln F^{(i)}(\tau^{(i)}-C^{(i)}x_{k})$  and $  - \ln ( 1 - F^{(i)}(\tau^{(i)}-C^{(i)}x_{k}) )$.
Since convexity is preserved under affine transformations, these terms are convex if and only if the functions $  \ell^{(i)}_1 (z) = - \ln F^{(i)} (z)$ and $\ell^{(i)}_2 (z) = - \ln (1- F^{(i)} (z) ) $ are convex.
A general proof of the convexity of these functions hinges on the observation that the functions $F^{(i)} (z) $ and $1-F^{(i)} (z)$ are obtained by integrating the log-concave
PDF of the measurement noise $v^{(i)}_k$ over the convex sets $(-\infty , z]$ and $(z, \infty)$, respectively.
Then, as discussed in previous work, \cite{Boyd} we can apply the integral property of log-concave functions \cite{} and show
that both $F^{(i)} (z) $ and $1-F^{(i)} (z)$ are log-concave, thus proving the convexity of  $  \ell^{(i)}_1 (z) $ and $\ell^{(i)}_2 (z) $.

Alternatively, a simple proof of the convexity of the functions $  \ell^{(i)}_1 (z) $ and $\ell^{(i)}_2 (z) $ can be given under the additional assumptions
that the PDF of the measurement noise is always strictly positive and differentiable. This implies that the PDF of the measurement noise can be written as
$e^{-\Phi^{(i)}(z) } $ for some convex differentiable function $\Phi^{(i)} (z)$.
We discuss in detail this case because it includes, among others, the normal distribution which is the most commonly used PDF
and because it provides useful insights on the form of the cost function.
In the following of the proof, to simplify the notation, the dependence on the sensor index $i$ will be dropped.
We start by noting that the CDF of the measurement noise can be written as
\[
F (z) = \int_{-\infty}^{z} e^{-\Phi (s)} ds \,
\]
Then, since the differentiability of $\Phi (z)$ implies that   $\ell_1 (z)$ and $\ell_1 (z)$ are twice differentiable in their domain,
convexity of these functions corresponds to the  second-order conditions $  \ell''_1 (z) \ge 0$ and  $ \ell''_2 (z) \ge 0$, respectively.
As it can be easily checked, we have
\begin{equation}\label{eq:logconc}
 \ell''_1 (z) = \frac{e^{-\Phi (z)}}{[F(z)]^2} \left [ e^{-\Phi (z)} + F(z) \, \Phi'(z)  \right ] \, .
\end{equation}
We distinguish two cases. When $ \Phi'(z)  \ge 0$, the second derivative is non-negative because all terms are non-negative.
When instead $ \Phi'(z)  < 0$, we can exploit the fact that, for a convex function $\Phi (z)$, we have $\Phi(s) \ge \Phi (z) + \Phi'(z) (s-z)$ to derive the upper bound
\[
F (z) = \int_{-\infty}^{z} e^{-\Phi (s)} ds \le e^{-\Phi (z)} e^{z \, \Phi' (z)}  \int_{-\infty}^{z} e^{-\Phi' (z) \, s} ds  = - \frac{e^{-\Phi (z)} }{\Phi' (z) } \, .
\]
Then, when $ \Phi'(z)  < 0$, we can lower bound the second derivative in (\ref{eq:logconc}) as
\[
  \ell''_1 (z) \ge \frac{e^{-\Phi (z)}}{[F(z)]^2} \left [ e^{-\Phi (z)} - \frac{e^{-\Phi (z)} }{\Phi' (z) } \, \Phi'(z)  \right ] = 0 \, ,
\]
proving the convexity of $\ell_1 (z) $. Similarly, for $\ell_2 (z)$ we have
\begin{equation}\label{eq:logconc2}
 \ell''_2 (z) = \frac{e^{-\Phi (z)}}{[1-F(z)]^2} \left [ e^{-\Phi (z)} - (1-F(z)) \, \Phi'(z)  \right ] \, .
\end{equation}
Again we distinguish two cases. When $ \Phi'(z)  \le 0$, the second derivative is non-negative because all terms are non-negative.
When instead $ \Phi'(z)  > 0$, we can exploit again the convexity of $\Phi (z)$ and derive the upper bound
\[
1- F(z) = \int_{z}^{\infty} e^{-\Phi (s)} ds \le e^{-\Phi (z)} e^{z \, \Phi' (z)}  \int_{z}^{\infty} e^{-\Phi' (z) \, s} ds  = \frac{e^{-\Phi (z)} }{\Phi' (z) } \, .
\]
Then, when $ \Phi'(z)  > 0$, we can lower bound the second derivative in (\ref{eq:logconc2}) as
\[
 \ell''_2 (z) \ge \frac{e^{-\Phi (z)}}{[1-F(z)]^2} \left [ e^{-\Phi (z)} - \frac{e^{-\Phi (z)} }{\Phi' (z) } \, \Phi'(z)  \right ] = 0 \, ,
\]
proving the convexity of $\ell_2 (z)$ and, hence, of the whole cost function.

\end{document}